\documentclass[twocolumn,showpacs]{revtex4}
\usepackage{graphicx}
\begin{document}
   \title{Propagation of light:\\Coherent  or Monte-Carlo computation ?}

   \author{J. Moret-Bailly}
   \email{jmo@laposte.net}
\date{\today}
 \begin{abstract}
 
   {Wrong Monte-Carlo computations are used to study the propagation of light in low pressure gas of nebulae.}
   {We recall that the incoherent interactions required for Monte Carlo calculations and hindering coherent interactions are due to collisions that disappear at low pressure. Incoherent interactions blur the images while coherent do not. We introduce coherent optical effects or substitute them for Monte Carlo calculations in published papers, improving the results and avoiding the introduction of ``new physics''.}
   {The spectral radiance of novae has the magnitude of the radiance of lasers, and large column densities are available in the nebulae. Several types of coherent interactions (superradiance, multiphoton effects, etc..), well studied using lasers, work in nebulae as in laboratories.}
   {The relatively thin shell of plasma containing atoms around a Str\"omgren sphere is superradiant, so that the limb of the sphere is seen as a circle which may be dotted into an even number of ``pearls``. The superradiant beams induce a multiphotonic scattering of the light rays emitted by the star, improving the brightness of the limb and darkening the star. Impulsive Stimulated Raman Scatterings (ISRS) in excited atomic hydrogen shift the frequencies of electromagnetic waves: UV-X lines of the Sun are red- or blue-shifted, the microwaves exchanged with the Pioneer 10 and 11 probes are blueshifted (no anomalous acceleration needed), the far stars are redshifted.}
  {Without any  ``new physics'', coherent spectroscopy works as a magic stick to explain  many observations.}
\end{abstract}

   \maketitle
   keywords: radiation mechanisms: nonthermal -- radiative transfers -- ISRS-- ISM:HII regions

\section{Introduction}
According to Einstein \cite{Einstein1917}, all interactions of light with matter are coherent in media large and homogeneous enough that statistical thermodynamics and Huygens construction are valid.
Most spectroscopists have long ignored Einstein's work. Thus they despised Townes' ideas before masers and lasers work. This skepticism persists in astrophysics where it also refers to an article in which Menzel \cite{Menzel} confused radiance and irradiance and wrote: {\it The so-called ``stimulated'' emissions which have here been neglected should be included where strict accuracy is required. It is easily proved, however, that they are unimportant in the nebulae.}  So, constantly, a large number of articles published in the best reviews, cited, for example, by Zheng \cite {Zheng}, Nilsson et al. \cite{Nilsson}, use Monte Carlo computations rather than the methods deducted from Einstein's theory, to calculate the propagation of light in atomic hydrogen.

Quantum mechanics associates a particle with a linear, ``Schr\"odinger wave'' $\Psi$, scalar field of complex value whose Hermitian square at a point is proportional to the probability of finding the particle at that point. It is not known how to define a $\Psi$ wave for the photon, so a scalar function of the electromagnetic field is used in its place. Associating photons to this wave requires the definition of ``normal modes" valid only in a limited optical system out of which the photons can not be exported. Thus, W.E. Lamb  \cite{WLamb}, W. E. Lamb, Schleich, Scully and Townes \cite{WLamb2} name the photon a pseudo particle which must be used with a great care that many physicists do not have. In the absence of Einstein's theory, astrophysicists apply Monte Carlo calculations to photons, in the study of light propagation in low pressure gas.

Section \ref{coherence} shows that the use of the Monte-Carlo computations in optics must be limited to the case of media so heterogeneous that the concept of light ray cannot be used.

Section \ref{Einstein} shows that the use of absolute spectral radiance simplifies the use of Einstein's theory in the presence of spontaneous emissions.

Section  \ref{Stromgren} improves a spectroscopic model introduced by Str\"omgren, in which many effects observed in laser technology apply.

Section \ref{Applications} justifies several theories done in astrophysics, theories rejected by authors whose spectroscopy did not take the coherence into account. It suggests other applications.

In conclusion, we suggest to develop models using optical coherence to simplify our sight of the universe.

\section{Coherence and incoherence.}\label{coherence}
The theory of light propagation in the atmosphere separates coherent and incoherent scatterings well:

Huygens' construction shows the propagation of monochromatic waves in a homogeneous continuous medium, considering that each infinitesimal fragment of the medium located on a surface wave emits a monochromatic wave of same frequency, coherent with the exciting wave. This construction also applies to particles of a real medium if the finite density of particles in the vicinity of a wave surface is large enough. It must be assumed that the molecules constituting the medium emit a wave having a well defined difference of phase with the local incident wave. In a transparent medium, this difference of phase is usually a delay of $\pi/2$. The identity of the initial and scattered wavefronts allows an interference of the two waves to produce a single monochromatic, refracted wave.

Einstein's theory \cite{Einstein1917}  extends the theory of refraction to the emission or absorption of light by assuming that the complex amplitude of an incident wave is multiplied by a complex ``amplification'' coefficient, preserving the wave surfaces and respecting the laws of thermodynamics. This extension is not obvious because the energy exchanges with the molecules are quantized, so we must admit the existence of a process of (de-) coherence, either quantum or classical by Huygens constructions. This process  transfers the energy exchanges between all involved molecules and few molecules that undergo a transition, and vice versa.

\medskip
Huygens' theory is faulty if the scattering of light by certain molecules depends on a stochastic parameter. In the neighborhood of a critical point of a gas, this parameter is the density that fluctuates and the theory of refraction is no longer valid. Away from the critical point, most molecules are no longer subject to density fluctuations, the coherent Rayleigh scattering (giving refraction) and incoherent Rayleigh scattering (blue of the sky ...) become compatible.

At low pressure, the fluctuations are mostly binary collisions, whose density is proportional to the square of the pressure: the incoherent scattering disappears at low pressure, for instance in the stratosphere. This conclusion, extended by Einstein's theory to all interactions, is exactly opposite to Menzel's statement.

\medskip
If Monte-Carlo calculations explain well the result of complex interactions (such as neutron and uranium atom, light and cloud droplets too inhomogeneous to form a rainbow), the draw of the phase of the pilot wave of the photon is a negation of all the wave aspect of light: In particular, two photons whose pilot waves have opposite phases must cancel and not add their effects. A Monte-Carlo calculation that would take all phases into account is very complicated and unnecessary as Huygens' construction provides the result of a large number of interferences.

\section{Using Einstein's theory.}\label{Einstein}
The formula for the spectral radiance of a blackbody at temperature $T_P$, given by Planck in 1900, was adapted in 1911 by its author to the absolute spectral radiance $I_\nu$ at frequency $\nu$ (Planck \cite{Planck1911}) and approved by Einstein and Stern \cite{Einstein1913}:
\begin{eqnarray}
I_\nu=\frac{h\nu^3}{c^2}\{1+\frac{2}{\exp(h\nu/kT_P)-1}\}.
\end {eqnarray}
A small hole in the body lets out a beam having this spectral radiance so that the formula defines the Planck temperature $T_P$ of any beam according to its absolute spectral radiance and its frequency.

Diffraction (and polarization) limited modes of interest in astrophysics propagate in beams limited by the aperture of a telescope and the diffracted image of a distant point. The optical extent (Clausius invariant) of these beams is the square of the wavelength $\lambda^2=c^2/\nu^2$. In these beams, an infinite, polarized sine wave defines a mode; each pulse of polarized natural light whose frequency spectrum $\Delta\nu$ corresponds to the pulse duration, defines a dynamical mode. Thus, the absolute energy in a mode is obtained multiplying the radiance $I_\nu$ by $\lambda^2/2=c^2/2\nu^2$. Corresponding to a degree of freedom, it is equivalent to $kT/2$ at high temperature, as required by thermodynamics. Using the absolute radiance, Einstein's A coefficient is null, so that there is no problem about the introduction of A: the spontaneous emission results from the amplification of the zero point field whose phase is generally unknown.

\medskip
Consider a transition between two nondegenerate levels of potential energy $E_1$ and $E_2>E_1$, whose respective populations are $N_1$  and $N_2$. Boltzman's law defines a transition temperature $T_B$ such that $N_2/N_1 =exp[(E_1-E_2)/kT_B]$. By convention, negative values of $T_B$ are accepted. In homogeneous environments, it is agreed to treat separately refraction and change of radiance. With this convention, a ray is attenuated or amplified by the medium which it crosses without its geometry is changed. Only the initial phase of a ray of unknown origin must be considered as stochastic. 

If the medium is opaque, $T_P$ equals $T_B$. Otherwise, $T_P$ tends to $T_B$. The algebraic value of the coefficient of amplification may be computed from Einstein's coefficient $B$, its sign is positive if $T_B<0$ or if $T_B>T_P$.

\medskip
An other advantage of the use of the absolute field is the correctness of the calculation of the energy exchanged with matter, which involves calculating the variation of the square of the absolute field which differs from the variation of the square of a relative field.

\section{Study of an astrophysical size model introduced by Str\"omgren.}\label{Stromgren}
\subsection{Str\"omgren's main results.}\label{results}
\begin{figure*}
\vspace*{0cm}
\includegraphics[width=10cm]{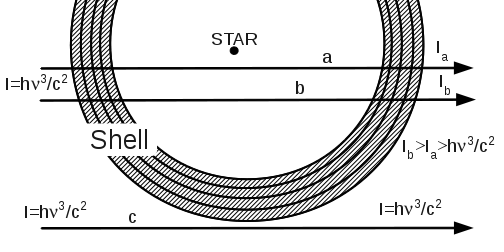}%
\caption{Comparison of the amplifications of rays crossing a Str\"omgren sphere: the path inside the infinitesimal shells is larger for ray b than for ray a, null for ray c. Thus, versus the distance of the ray from the star, the total amplification increases, then falls, therefore has at least a maximum.}
\label{coq}
\end{figure*}

Str\"omgren defined \cite{Stromgren} a model consisting of an extremely hot star immersed in a vast, low-density and initially cold hydrogen cloud. The ultra-violet emited by the star ionizes almost completely the hydrogen of a {\it Str\"omgren's sphere} which becomes transparent. Traces of atoms appearing in the outer regions of the sphere absorb energy by collisions and from light emitted by the star at their own eigen-frequencies.They dissipate this energy by radiating spontaneously into all directions.

\begin{figure*}
\vspace*{0cm}
\includegraphics[width=80mm]{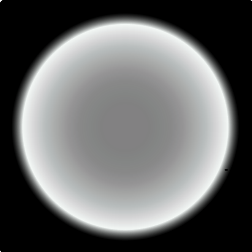}
\caption{Appearance of a Str\"omgren's shell, supposing that spontaneously emitted light is neither absorbed nor amplified.}
\label{sphere}
\end{figure*}

 This energy dissipation lowers the temperature and causes a catastrophic increase in the density of atoms, so that the sphere is surrounded by a {\it Str\"omgren's shell} relatively thin, radiating intensely.

\begin{figure*}
\vspace*{0cm} 
\vspace*{0cm} 
\includegraphics[width=80mm]{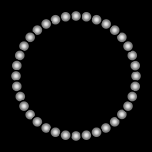}
\caption{With a strong superradiance a Str\"omgren's shell appears as a pearls necklace.}
\label{modes}
\end{figure*}

\subsection{Superradiance.}\label{superradiance}
 Figure \ref{coq} shows the amplification of a light ray by the Str\"omgren's shell which may be split into infinitesimal, concentric, spherical shells: For rays having crossed near the star, the path inside a crossed shell varies little. Farther, this thickness grows faster and faster up to a maximum amplification. Finally, the number of crossed infinitesimal shells decreases, the amplification falls to reach no amplification. Figure \ref{sphere} shows the appearance of the system supposing that the spontaneously emitted light is nether absorbed nor amplified.

Str\"omgren did not know the intense, coherent interactions of light with matter.
The plasma in the shell is similar to the plasma in gas lasers: the amplification factor is high, resulting in intense superradiance, maximal for rays located at a distance $R$ from the center of the sphere. $R$ defines precisely the inner radius of the shell. By competition of modes, these rays emit most of the available energy. Into a given direction they generate a circular cylinder.

If the superradiance is large, the competition of modes playing on these rays selects a system of an even number of orthogonal modes, as in a laser whose central modes have been extinguished, for example by an opaque disc (daisy modes). Thus, strongly superradiant rays of a transition form a circle regularly punctuated (Fig 3), and the atoms are quickly de-excited.
With a lower superradiance, or by a mixture of several frequencies in a wide band detector, the ring may appear continuous.

\subsection{Multiphotonic scattering}\label{multiphoton}
\medskip
The spectral radiance of the rays emitted by a supernova has, at each frequency, the radiance of a laser, so that multiphoton interactions and transitions between virtual levels are allowed. All frequencies may be involved in simple combinations of frequencies which result in resonance frequencies of hydrogen atoms. Thus, an important fraction of the energy emitted by the star is absorbed. All multiphoton absorptions and induced emissions form a parametric induced scattering that transfers the bulk of the energy of the radial rays emitted by the star to the superradiant rays. If the star is seen under a solid angle much smaller than the points forming the ring, it is no longer visible.

\subsection{Frequency shifts.}\label{shifts}
A large coherent transfer of energy between two co-linear light beams is difficult because the wavelengths being generally different, the difference of phase between the two beams changes, so that the transferred amplitudes cancel along the path. To preserve the coherence, the usual solutions are: the use of two indices of refraction of a crystal, the use of non-colinear broad light beams, or the use of light pulses:

In the ``Impulsive Stimulated Raman Scattering'' (ISRS) \cite{Ruhman}, the use of short laser pulses limits the phase shift of two light beams of slightly different frequencies, so that their coherence may be preserved: G. L. Lamb \cite{GLamb} wrote that the length of the used light pulses must be ``shorter than all relevant time constants''. A time constant is the collisional time because collisions break the phases. Others time constants are involved Raman periods which produce periods of beats between the exciting and the scattered beams.
 
\medskip
In appendix A, a general description of ISRS is avoided because the studied ``Coherent Raman Effect on Incoherent Light'' (CREIL) may be simply obtained by a replacement of the Rayleigh coherent scattering which produces the refraction by a Raman coherent scattering. For a ``parametric effect'', that is to avoid an excitation of matter so that matter plays the role of a catalyst, several light beams must be involved simultaneously. With the exception of frequency shifts, the properties of the CREIL are those of refraction:

- The interactions are coherent, that is the wave surfaces are preserved.

- The interaction is linear versus the amplitude of each light beam so that there is no threshold of amplitude.

- The entropy of the set of light beams increases by an exchange of energy which shifts their frequencies.

- Locally, the frequency shift of a beam is proportional to the column density of active scattering molecules and it depends on the temperatures and irradiances at all involved light frequencies.

- Lamb's conditions must be met. 

\medskip
Using ordinary incoherent light made of around one nanosecond pulses, the pressure of the gas and a Raman resonance frequency must be low, so that the effect is weak. With atomic hydrogen, frequency 1420 MHz of the hyperfine transition in 1S state is too large; in the first excited state the frequencies 178 MHz in the 2S$_{1/2}$ state, 59 MHz in 2P$_{1/2}$ state, and 24 MHz in 2P$_{3/2}$ are as large as allowed, very convenient.

\medskip
The background radiation is always implied. As it is generally nearly isotropic, a lot of beams is implied, their irradiance is large. Thus, for light, there is always a redshift component.
 
These frequency shifts are easily observed in laboratories using picosecond laser pulses, or, with longer pulses in optical fibers. The frequency shift is roughly inversely proportional to cube of the length of the pulses, so that astronomical paths are needed with usual light.

\begin{figure*}
\vspace*{0cm}
\includegraphics[width=100mm]{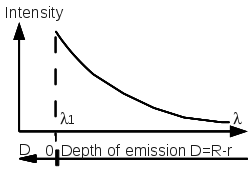}
\caption{Theoretical, qualitative spectrum of light spontaneously emitted along a ray crossing a Str\"omgren sphere. Observed at the surface of the sphere (D=0), the maximum of radiance is at the laboratory wavelength ($\lambda_1=\lambda_0$). In the Str\"omgren shell, all wavelengths decrease, the scale of the spectrum is changed: $\lambda_1<\lambda_0$. The spectrum depends on the distance $\rho$ between the ray and the star.}
\label{caj}
\end{figure*}

\medskip
The density of atomic hydrogen is negligible in Str\"omgren's sphere except near the surface, where it grows with an exponential look to the surface. Thus, the intensity of ``spontaneous emission'' is low in depth, high in surface. In propagating in a medium that contains excited hydrogen atoms, the beam provides energy to thermal radiation of high irradiance, and receives energy from the hot rays emitted by the star whose irradiance is low. It is reasonable assuming that the balance is negative, so that, at the surface, the weak, deep emission is redshifted, while the stronger, surface emission is at the laboratory wavelength $\lambda_0$  (Fig. \ref{caj} with $\lambda_1=\lambda_0 $, laboratory wavelength).

\medskip

In the shell, near the sphere, it remains excited hydrogen able to catalyze exchanges of energy. The energy emitted by the star which propagates radially at speed $c$ is transferred to the tangent rays whose {\it radial component of speed} is low, so that the irradiance of warm rays becomes very large: the cold spontaneous emission receives energy and the spectrum is shifted towards shorter wavelengths  (Fig. \ref{caj} with $\lambda_1<\lambda_0$).

\section{Possible applications in astrophysics.}\label{Applications}
\subsection{The distorted Str\"omgren's sphere of supernova remnant SNR1987A}

\begin{figure*}
\vspace*{0cm}
\includegraphics[width=150mm]{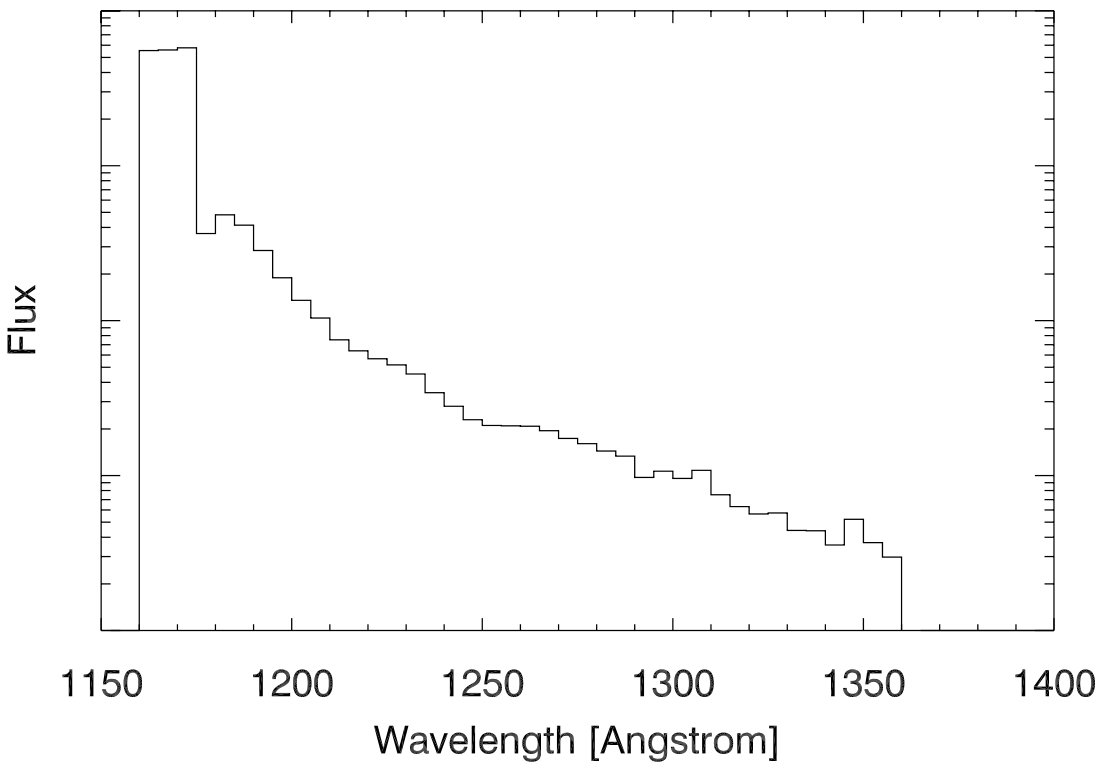}
\caption{Spectrum of the spontaneous emission of the disk inside the necklace of SNR1987A, computed by Michael et al. \cite{Michael}.}
\label{camic}
\end{figure*}

\begin{figure*}
\vspace*{0cm}
\includegraphics[width=150mm]{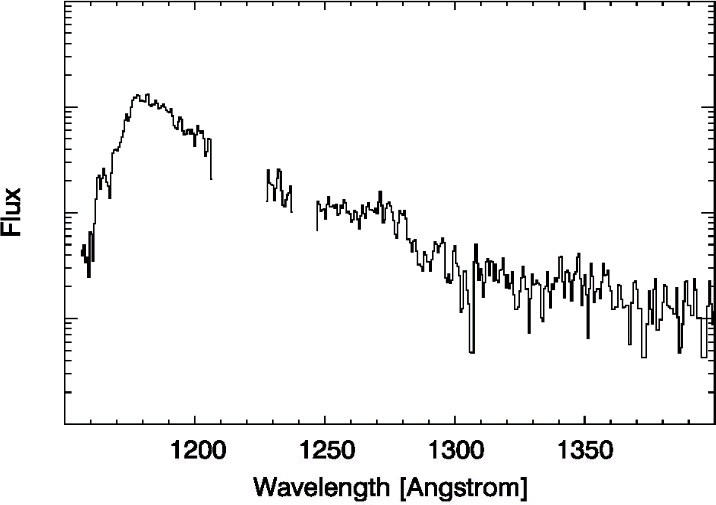}
\caption{Spectrum recorded inside the ring of SNR1987A.(Michael et al.\cite{Michael})} 
\label{recs}
\end{figure*}

The  supernova remnant SNR1987A is surrounded by relatively dense clouds of hydrogen making a ``hourglass'' (Sugerman et al.  \cite{Sugerman}). These clouds have been detected shortly after the explosion by photon echoes. Burrows et al. \cite{Burrows} criticized an interpretation of the three rings of SNR1987A as the limbs of the hourglass because, without superradiance, assuming that the hourglass is a strangulated Str\"omgren's sphere, gives a distorted figure similar to figure \ref{sphere}, not a distorted figure \ref{modes} as observed. The superradiance generates the three ``pearl necklaces'' at the limbs of the hourglass. Evidently it is necessary to take into account several lines of hydrogen, variations of the density of hydrogen, so that the monochrome image of SNR1987A is complicated.
Many images of nebulae, (for instance the ``bubble nebula'' SNR0509) are intermediate between figures \ref{sphere} and \ref{modes}. Without filters, SN1006 shows a sphere, its hydrogen lines show only a very bright part of its limb.

\medskip
The spectrum (Fig. \ref{recs}) emitted within the central ring of SNR1987A results from the superposition of spectra observed at different distances $\rho$ from the center of the disk, so that different spectra corresponding to different paths are added.

Our spectrum (Fig. \ref{caj}) does not show a strong, cut peak as the calculated spectrum by Michael et al. \cite{Michael} (Fig. \ref{camic}). Our computation is better than Monte-Carlo, but the starting point is the same: the redshift of the spectrum is assigned to the propagation of light in the plasma of hydrogen ra%\includegraphics{myfig.pdf}
ther than an expansion of the universe, a Doppler effect of winds,... . Probably Michael et al. did not insist on this point because of hostility against any discussion on the origin of the ``cosmological'' redshifts.

\subsection{Overview of other possible applications.}
Many ``planetary nebulae'' show arcs of circles or ellipses punctuated or not. The usual explanation is that the image of a very bright, distant star is distorted, multiplied by the gravitational lensing of an interposed, dark, heavy star. This explanation has been criticized because it involves a large number of alignment of proper stars with Earth, and because it is difficult to justify a certain regularity of punctuation. The phases of two contiguous dots are opposite, so that this feature can be tested by interference if the necklace is incomplete. Some, like ``the necklace nebula\textquotedblright{} are so similar to SNR1987A that they appear images of Str\"omgren's systems. The spectra could distinguish the two types of rings: the spectrum of the limb of a Str\"omgren's sphere is a line spectrum while the spectrum of a very far, bright star is probably a continuous emission spectrum.

Observed lines of many atoms have the shape of figure \ref{caj}. They may be generated by atoms heated in an hydrogen plasma.

\medskip
The frequency shifts by CREIL effect have many applications (Moret-Bailly  \cite{MBIE,MB0507141,MBAIP06}:

 - Increase in the frequency of the radio-waves exchanged with the Pioneer 10 and 11 probes, resulting from a transfer of energy from the solar radiation where the solar wind is cooled enough to generate atoms. This frequency shift is usually interpreted by Doppler effect, as an anomalous acceleration of the probes.

 - The frequency shifts of the extreme UV lines emitted by the Sun and observed by SOHO are usually attributed to a Doppler effect due to vertical speeds of the source. But the frequencies observed at the limb are not the laboratory frequencies. Assume that at high pressure and temperature, hydrogen is in a state simil%\includegraphics{myfig.pdf}
ar to a crystal, so that a CREIL effect is possible. The paths, thus the frequency shifts, from the depth that emits a line are larger at the limb than at the center so that the laboratory frequencies are preserved.

 - The spectrum of a neutron star heated to a very high temperature by accretion of a cloud of hydrogen is very similar to the spectrum of a quasar, including the Karlsson's periodicities. Thus the quasars may be in our galaxy or close to it, so that they are not enormous and do not move very fast.

 - High redshifts appear where hydrogen is  atomic and excited.

 - It is necessary to re-examine the scales of distances deduced from Hubble's law which assumes a redshift of the spectra proportional to the path of light whereas the example of SNR1987A shows that it is necessary to take account of other parameters: density and state of hydrogen, temperature of the studied rays, temperatures and radiances of the other rays. Thus, an important work appears necessary to press the sponge of the maps of the galaxies. To get some reliable distances, one can use the dynamics of the galaxies to evaluate, without black matter, their sizes thus their distances.

\section{Conclusion}
The introduction of optical coherence in astrophysics provides new, efficient tools able to deepen our understanding of the universe:

\medskip
The propagation of light in resonant, diluted gases is usually calculated by two methods: optical coherence (Einstein) or Monte Carlo calculations. The results are very different so that Einstein's theory largely verified by the success of gas lasers must be chosen. The Monte Carlo calculations should be reserved for the propagation of light in opalescent media.

The power of tools developed in connection with lasers should be used to study the diluted gases present in interstellar space with column densities much higher than those typically found in gas lasers.

Many observations are easily explained by optical coherence: Papers whose conclusions are not convincing are validated by introducing coherence: The ``pearls necklaces" and the multiple images of stars attributed to gravitational lensing, arise from superradiance.

The optical coherence explains the disappearance of supernova 1987A when its ``pearls necklace" appeared. The superradiance validates their coincidence with the limbs of an ``hourglass" observed by photon echoes: it sharpens and ponctuates this limb.

The shape of many spectral lines broken at the shortest wavelength is due to a spontaneous emission and redshift of the lines in a hydrogen plasma.

\medskip
Unexpected results occur:

  - The frequencies of UV-X spectral lines of the Sun coincide with laboratory frequencies assuming that the lines are not shifted by a Doppler effect, but by an  ``Impulsive Stimulated Raman Scattering" (ISRS): Energy is exchanged between radiations propagating in hot compressed hydrogen similar to a crystal.

  - The ``anomalous acceleration" of Pioneer 10 and 11 results from the attribution to a Doppler effect, of the blueshift of the carrier of microwave signals. This shift is due to an exchange of energy between the solar light and the microwaves.

- The Hubble law can be explained by an exchange of energy between light and the microwave background by ISRS. This law does not provide a reliable distance scale as the ISRS depends, in particular, on the density of excited atomic hydrogen that works as a catalyst.

\appendix
\section{Appendix: Frequency shifts of time-incoherent light beams by coherent transfers of energy (CREIL).}
The CREIL is a parametric effect resulting from the assembly of several Impulsive Stimulated Raman Scatterings, and applied to temporally incoherent radiation. Here, it is deduced from refraction, replacing the Rayleigh coherent scattering by Raman scatterings.

To explain the wave propagation, Huygens deduced from a wave surface known at time $t$, a slightly later wave surface, at time $t+\Delta t$. For that, he supposes that each element of volume contained between wave surfaces relating to times $t-\mathrm dt$ and $t$, emits at the local speed of the waves $c$, a spherical wavelet of radius $c\Delta t$. The set of these wavelets has, as envelopes, the sought wave surface and a retrograde wave surface canceled by the retrograde waves emitted at other times.

Let us suppose that each element of volume considered by Huygens contains molecules which emit also a wavelet of much lower amplitude at same frequency, whose phase is delayed by $\pi/2$ (Rayleigh coherent emission). Both types of emission generate the same wave surfaces, so that their emitted fields may be simply added. Are $E_0 \sin (\Omega T)$ the\label{shell} incident field, $E_0K\epsilon \cos(\Omega T)$ the field diffused in a layer of infinitesimal thickness $\epsilon=c\mathrm dt$ on a wave surface, and $K$ a coefficient of diffusion. The total field is:
\begin{equation}
E=E_0[\sin(\Omega t)+K\ % \date{Received September 15, 1996; accepted March 16, 1997}
epsilon \cos(\Omega t)]\label{refr}
\end{equation}
\begin{equation}
\approx E_0[\sin(\Omega t)\cos(K\epsilon)+\sin(K\epsilon )\cos(\Omega t)]=E_0\sin(\Omega t -K\epsilon).
\end{equation}

 This result defines the index of refraction $n$ by the identification 
\begin{equation}
K=2\pi n/\lambda=\Omega n/c.\label{index}
\end{equation}

Try to replace in this theory of refraction the Rayleigh scattering by a Raman scattering, with a shifting frequency $\omega$, but no initial phase shift.

Setting $K'>0$ the anti-Stokes diffusion coefficient, formula \ref{refr} becomes:

\begin{equation}  
E=E_0[(1-K'\epsilon)\sin(\Omega t)+K'\epsilon \sin((\Omega+\omega)t)].
\end{equation}

In this equation, incident amplitude is reduced to obtain the balance of energy for $\omega=0$.

\begin{eqnarray}
E=E_0\{(1-K'\epsilon)\sin(\Omega t)+\nonumber\\
+K'\epsilon[\sin(\Omega t)\cos(\omega t)+\sin(\omega t)\cos(\Omega t)]\}.
\end{eqnarray}

$K'\epsilon$ is infinitesimal; suppose that between the beginning of a pulse at $t=0$ and its end, $\omega t$ is small; the second term cancels with the third, and the last one transforms:

\begin{eqnarray}
E\approx E_0[\sin\Omega t+\sin(K'\epsilon\omega t)\cos(\Omega t)]\nonumber\\
E\approx E_0[\sin(\Omega t)\cos(K'\epsilon\omega t)+ \sin(K'\epsilon\omega t)\cos(\Omega t)\\
E\approx E_0\sin[(\Omega+K'\epsilon\omega)t].\label{eq4}
\end{eqnarray}

Hypothesis $\omega t$ small requires that Raman period $2\pi/\omega$ is large in comparison with the duration of the light pulses; to avoid large perturbations by collisions, the collisional time must be larger than this duration.  This is a particular case of the condition of space coherence and constructive interference written by Lamb.

Stokes contribution, obtained replacing $K'$ by a negative $K''$, must be added. Assuming that the gas is at temperature $T$, $K'+K''$ is proportional to the difference of populations in Raman levels, that is to $\exp[-h\omega/(2\pi kT)]-1 \propto \omega/T$.

$K'$ and $K''$ obey a relation similar to relation \ref{index}, where Raman polarisability which replaces the index of refraction is also proportional to the pressure of the gas $P$ and does not depend much on the frequency if the atoms are far from resonances; thus, $K'$ and $K''$ are proportional to $P\Omega$, and $(K'+K'')$ to $P\Omega \omega/T$. Therefore, for a given medium, the frequency shift is:
\begin{equation}
\Delta\Omega=(K'+K'')\epsilon\omega\propto P\epsilon\Omega\omega^2/T.\label{delom}
\label{redshift}
\end{equation}
The relative frequency shift $\Delta\Omega/\Omega$ of this space-Coherent Raman Effect on time-Incoherent Light (CREIL) is nearly independent on $\Omega$ and proportional to the integral of $Pc{\rm d}t$, that is to the column density of active gas along the path.

The path needed for a given (observable) redshift is inversely proportional to $P\omega^2$. At a given temperature, assuming that the polarisability does not depend on the frequency, and that $P$ and $\omega$ may be chosen as large as allowed by Lamb's condition, this path is inversely proportional to the cube of the length of the pulses: an observation easy in a laboratory with femtosecond pulses requires astronomical paths with ordinary incoherent light.

\bibliographystyle{unsrt}%plain
\bibliography{JMB}
\end{document}